\documentclass{80SB}
\usepackage{txfonts}
\usepackage{graphicx}

\title{Evolution of the spin resonance of CeCoIn$_{5}$ as a function of magnetic field and La substitution}

\author{St\'ephane Raymond, Justin Panarin, G\'erard Lapertot and Jacques Flouquet} 

\inst{SPSMS, UMR-E 9001, CEA-INAC/UJF-Grenoble 1, 38054 Grenoble, France}

\abst{We report the evolution of the spin resonance in CeCoIn$_{5}$ as a function of magnetic field and lanthanum substitution. In both cases, the resonance peak position shifts to lower energy and the lineshape broadens. For La doping, it is found that the ratio $\Omega_{res}/k_{B}T_{c}$ is almost constant as a function of $x$. Under magnetic field the decrease of the excitation energy is similar for H// [1,$\bar{1}$,0] and [1,1,1] and faster than the decrease of $T_{c}(H)$. The Zeeman effect found for the field applied along [1,$\bar{1}$,0] corresponds to the ground state magnetic moment.}

\kword{Unconventional superconductivity, heavy fermion systems, inelastic neutron scattering}

\begin{document}
\maketitle

\section{Introduction} 
CeCoIn$_{5}$ has the highest superconducting transition temperature among cerium heavy fermion materials ($T_{c}$ = 2.3 K) \cite{Petrovic}. It belongs to the so called 1-1-5 family of compounds (CeTIn$_{5}$, T=Co, Rh, Ir) where the competition between magnetism and superconductivity can be finely tuned by applying pressure, chemical substitution or magnetic field leading to rich interpenetrated magnetic and superconducting phase diagrams \cite{Sarrao,Knebel}. CeCoIn$_{5}$ is identified as a $d$-wave superconductor \cite{An} with multiband effects \cite{Seyfarth}. It crystallizes in the tetragonal space group P4/mmm and the Fermi surface shows two dimensional features \cite{Fermi}. The closeness of CeCoIn$_{5}$ to a magnetic quantum critical point is attested by measurements for various doping \cite{Zaum} and under magnetic field \cite{Ludovic}. One of the most important aspect of the physics of CeCoIn$_{5}$ is a new phase that appears at high field and low temperature (HFLT phase, H$\ge$ 10.5 T, T$\le$ 350 mK) when the field is applied in the basal plane of the tetragonal structure. Initially this state was considered\cite{Bianchi} as the realization of a modulated superconducting FFLO state \cite{theory}. In contrast, up to now, the only microscopic signature of the HFLT phase is a long range magnetic order of incommensurate nature with the magnetic moments along the $c$-axis and a moment of about 0.15 $\mu_{B}$\cite{Kenzelmann1,Kenzelmann2,Blackburn}. Strikingly this magnetic order is tight to the superconductivity and disappears above $H_{c2}$ and this unusual feature leads to many theoretical propositions \cite{Yanase,Ikeda,Kato}.

\section{Resonance as a function of magnetic field}
As concerns the magnetic excitation spectrum at zero magnetic field, a spin resonance is observed below $T_{c}$ and is peaked at the hot-spot vector $\textbf{Q}$=(1/2, 1/2, 1/2) with an energy $\Omega_{res}$ = 0.55 meV \cite{Stock}. We previously reported the evolution of the spin resonance of CeCoIn$_{5}$ for magnetic field applied along [1,$\bar{1}$,0] direction \cite{Panarin1} and for La substitution \cite{Panarin2}. 

In the present paper, we show new data for magnetic field applied along [1,1,1]. This latter experiment was performed on the cold neutron three axis spectrometer IN12 at ILL, Grenoble. Measurements were performed using the 3.8 T horizontal field magnet with a dilution insert.  The incident beam was provided by a vertically focusing pyrolytic graphite (PG) monochromator. A Be filter was placed just before the sample in order to cut down the higher order contamination of neutrons. The sample is the same as in Panarin et al.\cite{Panarin1} and the [1,1,0] and [0,0,1] directions define the horizontal scattering plane. The spectrometer was setup in W configuration using constant $k_{f}$=1.3 $\AA^{-1}$, a horizontally focusing PG analyzer was used with collimations 60'-open-open. The data analysis is the same as already reported in the previous papers \cite{Panarin1,Panarin2}. Figure 1 shows spectra measured for different magnetic fields applied along $\textbf{Q}$=(1/2, 1/2, 1/2)  at 100 mK. The resonance peak position shifts to lower energy when the magnetic field increases while the peak lineshape broadens. This behavior is very similar to the one reported for H//[1,$\bar{1}$,0] \cite{Panarin1}.
\begin{figure}
%\vspace{0.7cm}
\centering
\includegraphics[width=7cm]{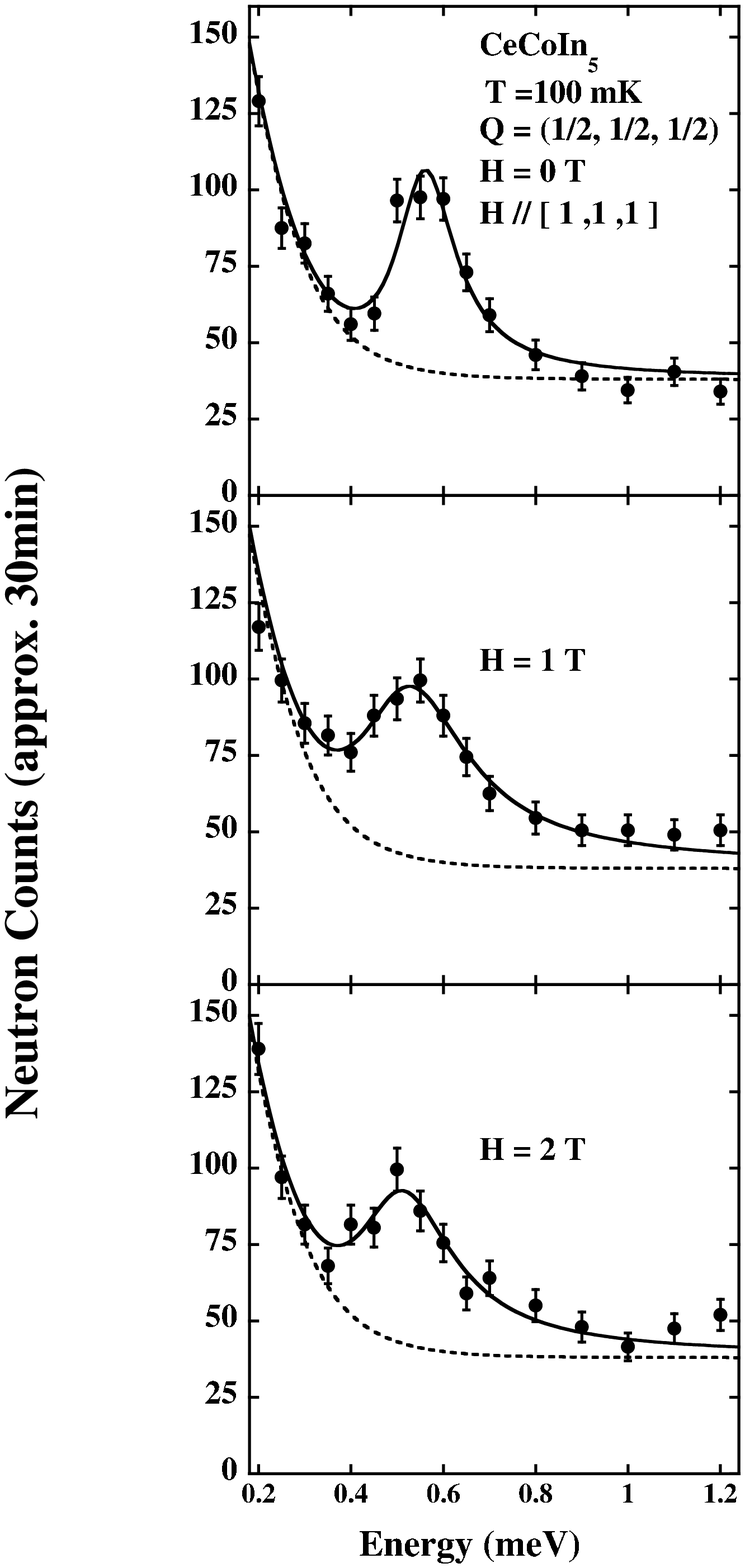}
%\vspace{ -1.4cm}
\caption{Excitation spectrum of CeCoIn$_{5}$ measured at $\bf{Q}$=(1/2, 1/2, 1/2) for $H$ = 0,  1 and  2 T applied along [1,1,1] and $T$=100 mK. The solid lines are inelastic Lorentzian fits as described in Panarin et al.\cite{Panarin1,Panarin2}. The dashed line indicates the background.}
\end{figure}
Figure 2 shows the field variation of the peak position as a function of magnetic field applied along [1,$\bar{1}$,0] and [1,1,1]. The latter data fall on the same curve than the one along [1,$\bar{1}$,0]. A linear fit gives a slope of $\alpha$=-0.039(2) meV.T$^{-1}$. The corresponding linear extrapolation to zero energy  of the resonance peak will give a critical field of 14.1 T.

\begin{figure}[t!]
%\vspace{-2.5cm}
\centering
\includegraphics[width=7cm]{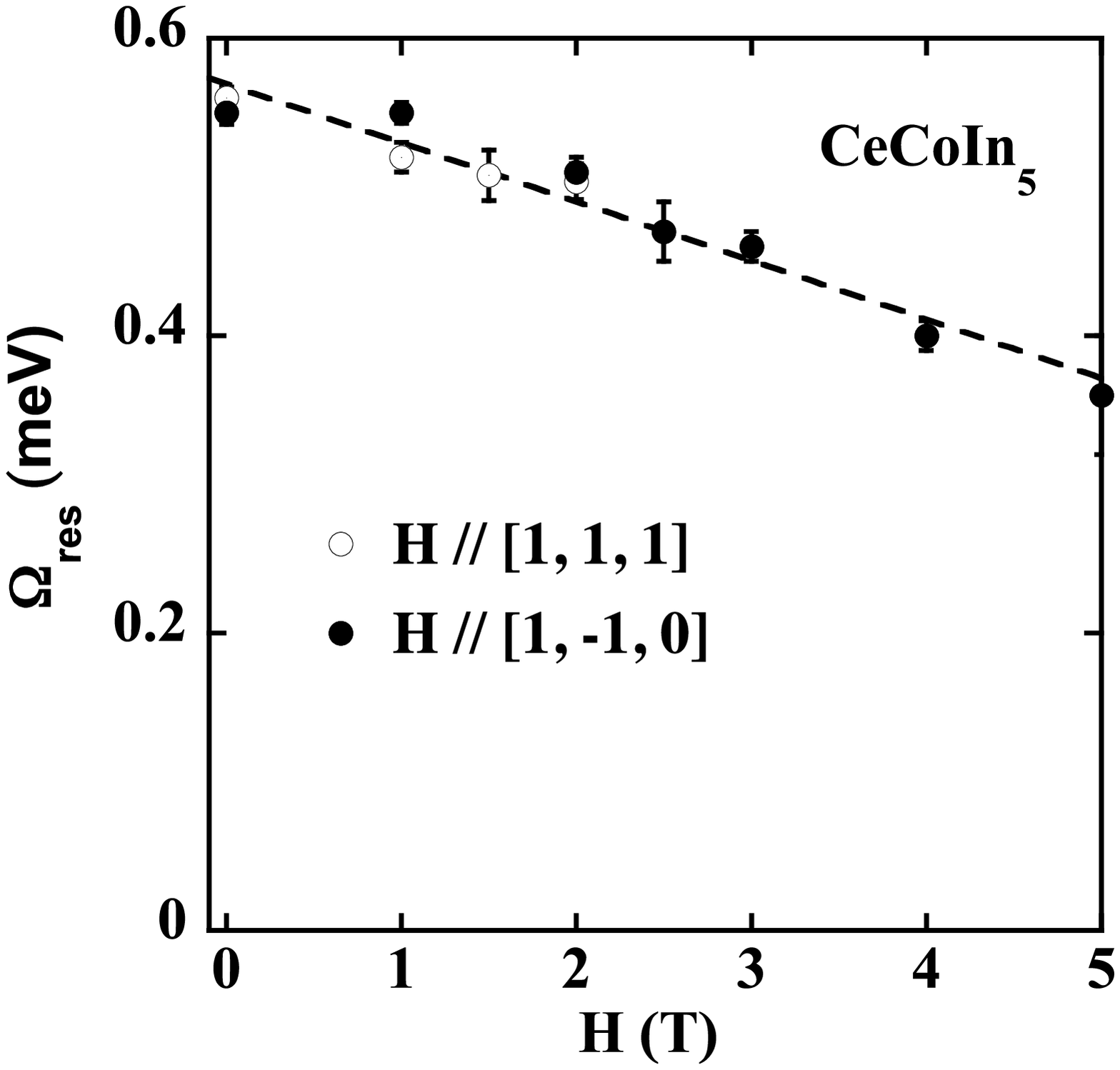}
%\vspace{ -2cm}
\caption{Magnetic field dependence of the resonance energy for magnetic field applied along [1,$\bar{1}$,0] and [1,1,1].}
\end{figure}

\section{Analysis of the Zeeman effect}
In line with the works performed on cuprates, the nature of the resonance peak in CeCoIn$_{5}$ is in debate being described either as an exciton (a S=1 bound state in the particle-hole channel) \cite{Eremin} or as a magnon (a magnetic mode visible below $T_{c}$ due to the suppression of Landau damping) \cite{Chubu}.
Without precise theoretical model, it is therefore difficult to analyze our data beyond a phenomenological approach.
Since $\Omega_{res}$ is associated with $d$-wave superconductivity, one can expect that it will collapse at $H_{c2}$.  This behavior has been observed for the spin gap of the electron doped compound Nd$_{1.85}$Ce$_{0.15}$CuO$_{4}$ \cite{Motoyama}. In this view, it is suprising that the initial field variation of $\Omega_{res}$ found in CeCoIn$_{5}$ is very similar for both field directions shown in Fig.2 whereas the upper critical field $H_{c2}$ is 11.8 T in the plane  and respectively 9 T along [1,1,1] (this value is taken from Correa et al. \cite{Correa } knowing that the angle between the basal plane and [1,1,1] is 23$^{\circ}$ ; the given directions are in reciprocal space). If $\Omega_{res}$ must vanish at $H_{c2}$, one would expect that $\Omega_{res}$ will decrease strongly for the direction of the smaller critical field. However in low field, the difference in $T_{c}(H)$ between the basal plane and [1,1,1] is rather weak.

Beyond the fact that $\Omega_{res}$ is associated with superconductivity, it is at first place a magnetic excitation and its response to a magnetic field must be considered to this respect. The first unanswered question concerns the polarization of the resonance in relation with the magnetic anisotropy of the system. On the one hand, the bulk susceptibility shows that the $c$-axis is the easy axis by a factor of about two at low temperatures \cite{Petrovic}. On the other hand, recent NMR experiments\cite{NMR} point out that the staggered susceptibility has exactly the reverse anisotropy than the bulk one. Further neutron scattering experiments are needed to address this point. The second question concerns the degeneracy of the excitation that is also not clear at present. In the most common model of spin exciton, the mode is a triplet. While there is a consequent amount of work  dealing with the response of a triplet excitation under several magnetic field directions for insulating model magnets \cite{Regnault}, one must be careful when applying these ideas to complex metallic systems since both damping effects and cross-section issues may render some modes not measurable \cite{PrOs}. As a consequence we do not have at present a sound explanation for the similar field dependence of $\Omega_{res}$ for H //  [1,$\bar{1}$,0] and [1,1,1]. All the more the statistics of our data is limited and the two field directions are quite close. The most interesting experiment  would be to put the field along [0,0,1] but this is not technically  feasible while measuring at $\bf{Q}$=(1/2,1/2,1/2). 

In our measurements, we observed only one mode of decreasing energy when the magnetic field increases. One can consider the slope of the observed Zeeman effect. We calculate that the ground state wave function obtained from x-ray and neutron scattering experiment \cite{Willers} ($|0\rangle=0.36|\pm 5/2\rangle+0.93|\mp 3/2\rangle$) carries the (para)magnetic moment $\mu_{z}$=0.83 $\mu_{B}$ (along the $c$-axis) and $\mu_{perp}$=0.64 $\mu_{B}$ (in the plane). Knowing that $1 \mu_{B} \times$ 1 T $ \approx $ 0.058 meV, we conclude from the slope of $\Omega_{res}(H)$ that the magnetic mode carries the magnetic moment of the crystal field ground state (0.64 $\times$ 0.058 = 0.03712 $\approx$ $\alpha$). This striking feature would mean that the resonance is not a $S$=1 exciton as in the most common models for such an excitation. In other words, if we describe the ground state with a pseudo-spin 1/2, the measured excitation also carries the same pseudo-spin 1/2. Furthermore, we observe the Zeeman effect of the 4$f$ localized crystal field magnetic moment. This may have some relevance to theories pointing the importance of localized electrons in CeCoIn$_{5}$ \cite{Pines,Coleman}. However if this alone will control the field dependence of $\Omega_{res}$, a slight faster decrease is expected for field along [1,1,1].

\section{Resonance versus HFLT phase}
Finally let us comment on the relevance of the resonance peak for the HFLT phase. Instead of considering that $\Omega_{res}$ collapses at $H_{c2}$, another viewpoint is to consider that $\Omega_{res}$ collapses at $H_{HFLT}$ lower than $H_{c2}$. In this view, magnetic order occurs at $H_{HFLT}$ in analogy with Goldstone mode for a phase transition. This behavior has been observed for the spin gap of the hole doped compound La$_{1.855}$Sr$_{0.145}$CuO$_{4}$ \cite{Chang} where long range order occurs for magnetic field one order of magnitude below $H_{c2}$.This simple idea corresponds to an "old" prediction of SO(5) theory \cite{SO5}. The same viewpoint  is developed in a recent model of exciton condensation leading to long range magnetic order \cite{Michal}. Both elastic and inelastic neutron scattering data indicate that this situation occurs for H // [1,$\bar{1}$, 0]. In contrast, for H // [1,1,1], at a tilt angle of 23$^{\circ}$ from the basal plane, the magnetic order is not reported: it disappears at 17$^{\circ}$ according to neutron scattering data \cite{Blackburn} and at 18$^{\circ}$ according to bulk measurements \cite{Correa}. For this field direction, we observed a decrease of the resonance energy. Hence there is not a one to one correspondence between the response of the resonance under field and the occurrence of the HFLT phase : not only the mode must soften but also the susceptibility must be enhanced in order to favor magnetic order \cite{Michal}. The sensitivity of the HFLT phase to the magnetic field orientation means that stringent conditions are required for the cooperative effect between magnetism and superconductivity \cite{Ikeda,Kato,Michal}.
\begin{figure}[t!]
%\vspace{-2.5cm}
\centering
\includegraphics[width=7cm]{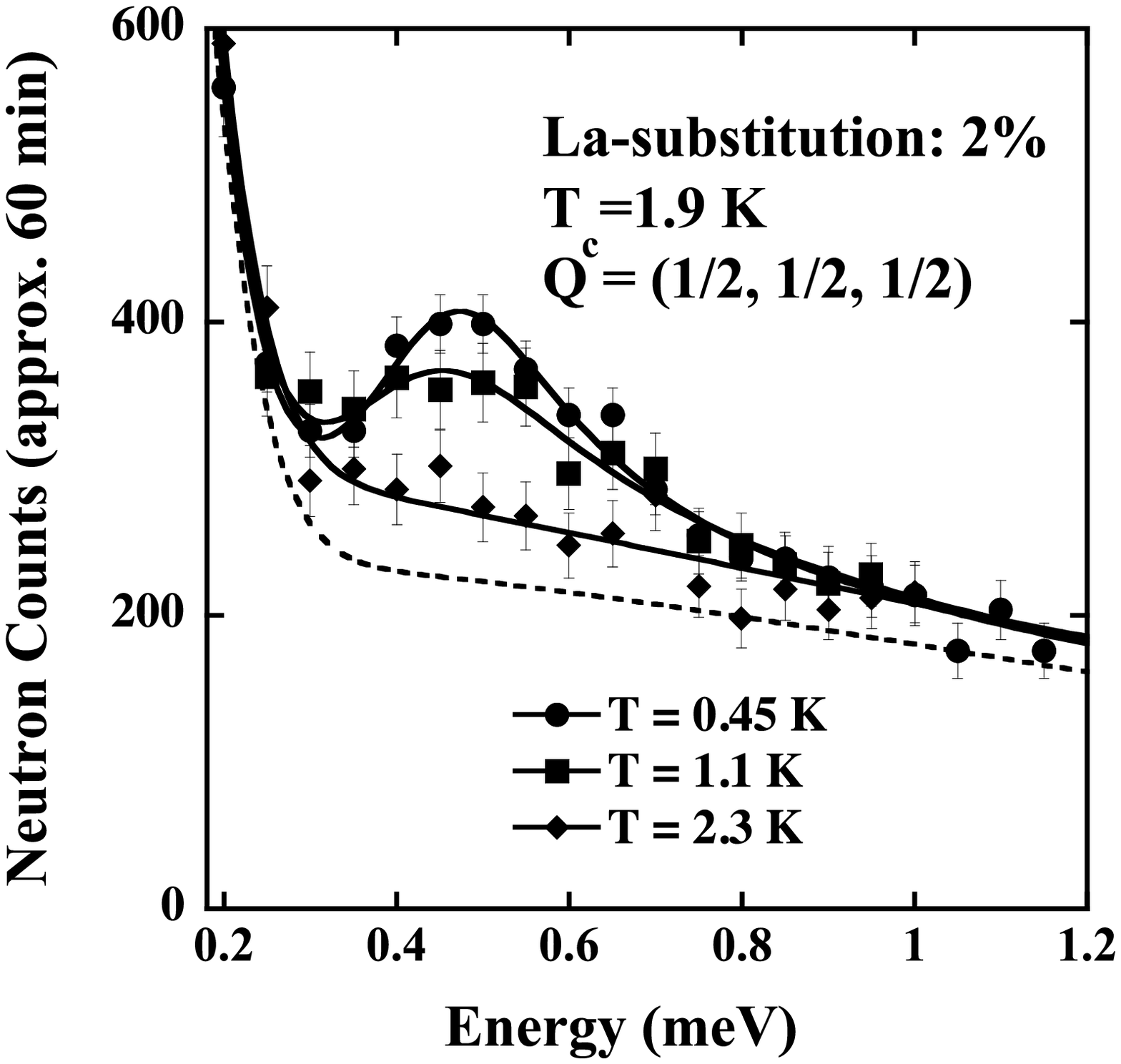}
%\vspace{ -2cm}
\caption{Temperature dependence of the magnetic excitation spectrum of Ce$_{0.98}$La$_{0.02}$CoIn$_{5}$ measured at $\textbf{Q}$=(1/2, 1/2, 1/2).}
\end{figure}
\section{Resonance as a function of La substitution}
The experiments carried out with La doping lead to the conclusion that $\Omega_{res}/k_{B}T_{c}$ is almost constant as a function of La concentration \cite{Panarin2}. Figure 3 shows the temperature dependence of the magnetic excitation spectrum for Ce$_{0.98}$La$_{0.02}$CoIn$_{5}$ for which $T_{c}$=1.9 K. As for the pure compound\cite{Stock}, the magnetic excitation does not exist above $T_{c}$ and the resonance at low energy peaks at $\Omega_{res}$=0.45 meV.
Figure 4 shows the variation of $\Omega_{res}/k_{B}T_{c}$ as a function of La concentration as reported in Panarin et al.\cite{Panarin2}.
As already stated the theory for this mode is not settled. The only theoretical calculation dealing with impurity effects on a spin resonance is performed in the framework of the spin exciton model for the cuprates \cite{Li}. It predicts the two effects observed here: decrease of the resonance energy and lineshape broadening. By examining the data on cuprates, iron superconductors and heavy fermion systems,  it was reported that the resonance energy is in a phenomenological way proportional to the superconducting gap with $\Omega_{res} \approx 0.64 \times 2 \Delta$ \cite{Yu}. Since for a $d$-wave superconductor, it is expected that pair-breaking non magnetic impurities lead to $\Delta/\Delta_{0}=T_{c}/T_{c0}$ \cite{Maki}, the relation $\Omega_{res} \propto k_{B}T_{c}$ is somehow expected in the case of La substitution ($\Delta_{0}$ is the gap for $x$=0, $T_{c0}$ is the value of $T_{c}$ for $x$=0). The fact that pair-breaking is the dominant effect in Ce$_{1-x}$La$_{x}$CoIn$_{5}$, as opposed to a "simple" tuning parameter is underlined by resistivity \cite{Petrovic2} and specific heat measurements \cite{Tanatar}. 
\begin{figure}[t!]
%\vspace{-2.5cm}
\centering
\includegraphics[width=7cm]{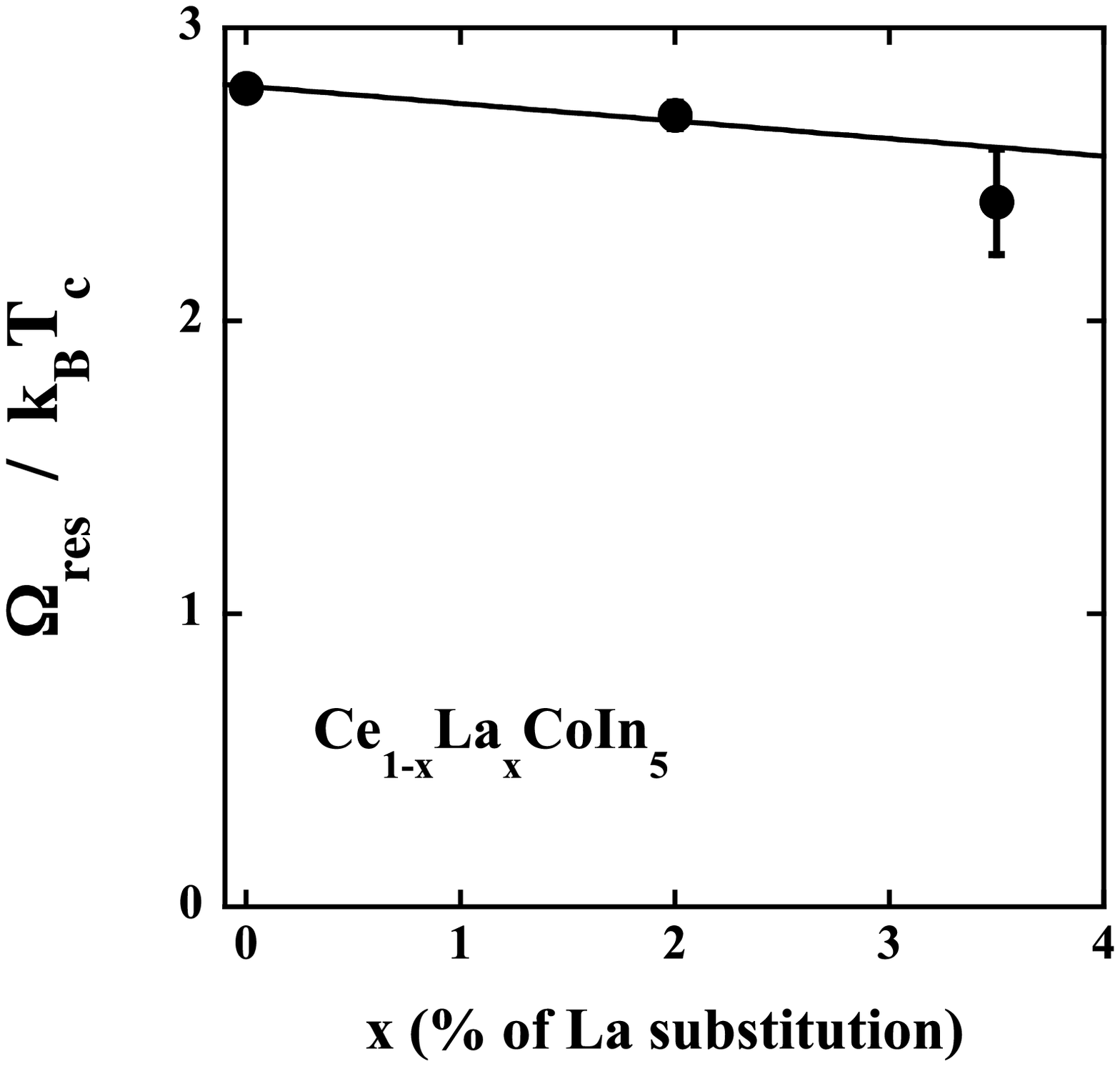}
%\vspace{ -2cm}
\caption{$\Omega_{res}/k_{B}T_{c}$ as a function of La concentration in Ce$_{1-x}$La$_{x}$CoIn$_{5}$.}
\end{figure}
\section{$\Omega_{res}(H)$ versus $T_{c}(H)$}
Under magnetic field, the situation is more complex, since in CeCoIn$_{5}$ both Pauli and orbital effects are present as the magnetic field increases. In the model of the spin exciton these effects are treated separately: the orbital effect is treated by Eschrig et al.\cite{Eschrig}: it leads to a decrease of $\Omega_{res}$ ;  the Pauli effect is treated by Ismer et al.,\cite{Ismer}: it leads to Zeeman splitting of the excitation. In the work of Michal et al.\cite{Michal}, the higher mode of the split peaks is damped by the continuum of excitations. Another difficulty is that the magnetic field variation of the superconducting gap $\Delta$ is not simple as compared to the case of impurities reported above. Contrasted behaviors for $\Delta(H)$ are found in the literature \cite{Vedeneev}.  In the line of the work performed on La substitution,  we use  $T_{c}(H)$ as a normalization factor, which has no justification a priori. It is nevertheless useful in order to compare the field variation of $\Omega_{res}(H)$ and $T_{c}(H)$.  In figure 5, we plot $\Omega_{res} \times T_{c0}/ T_{c}$ ($T_{c0}$ is $T_{c}$ for $H$=0 T) as a function of $H/H_{c2}$ for CeCoIn$_{5}$ ($H$ // [1, $\bar{1}$, 0]) and UPd$_{2}$Al$_{3}$\cite{Blackburn2} ($H$ // $b$). In CeCoIn$_{5}$, this quantity decreases and eventually will vanish around $Hc_{2}$ or $H_{HFLT}$ (see discussion above). In contrast, this quantity is almost constant in UPd$_{2}$Al$_{3}$, which means that in this compound $\Omega_{res}(H) \propto T_{c}(H)$. Both orbital and Pauli effects are present in UPd$_{2}$Al$_{3}$ \cite{Hc2}, so that such a simple behavior is not easy to interpret. It is worthwhile to note that in UPd$_{2}$Al$_{3}$, the superconductivity occurs inside an antiferromagnetic phase and this complicated the interpretation of the resonance mode behavior under field \cite{Note}. Among the few unconventional superconductors for which a careful magnetic field study of the resonance/spin gap mode is available, we notice that when magnetic field induces antiferromagnetism (CeCoIn$_{5}$, La$_{1.855}$Sr$_{0.145}$CuO$_{4}$), $\Omega_{res}(H)$ drops faster than $T_{c}(H)$. This would give credit to the soft mode mechanism for the magnetic ordering discussed above. In contrast, when the magnetic field does not displace the boundary between magnetism and superconductivity (beyond upper critical field effect), it is found that $\Omega_{res}(H) \propto T_{c}(H)$ (UPd$_{2}$Al$_{3}$, Nd$_{1.85}$Ce$_{0.15}$CuO$_{4}$).
\begin{figure}[t!]
%\vspace{-2.5cm}
\centering
\includegraphics[width=7cm]{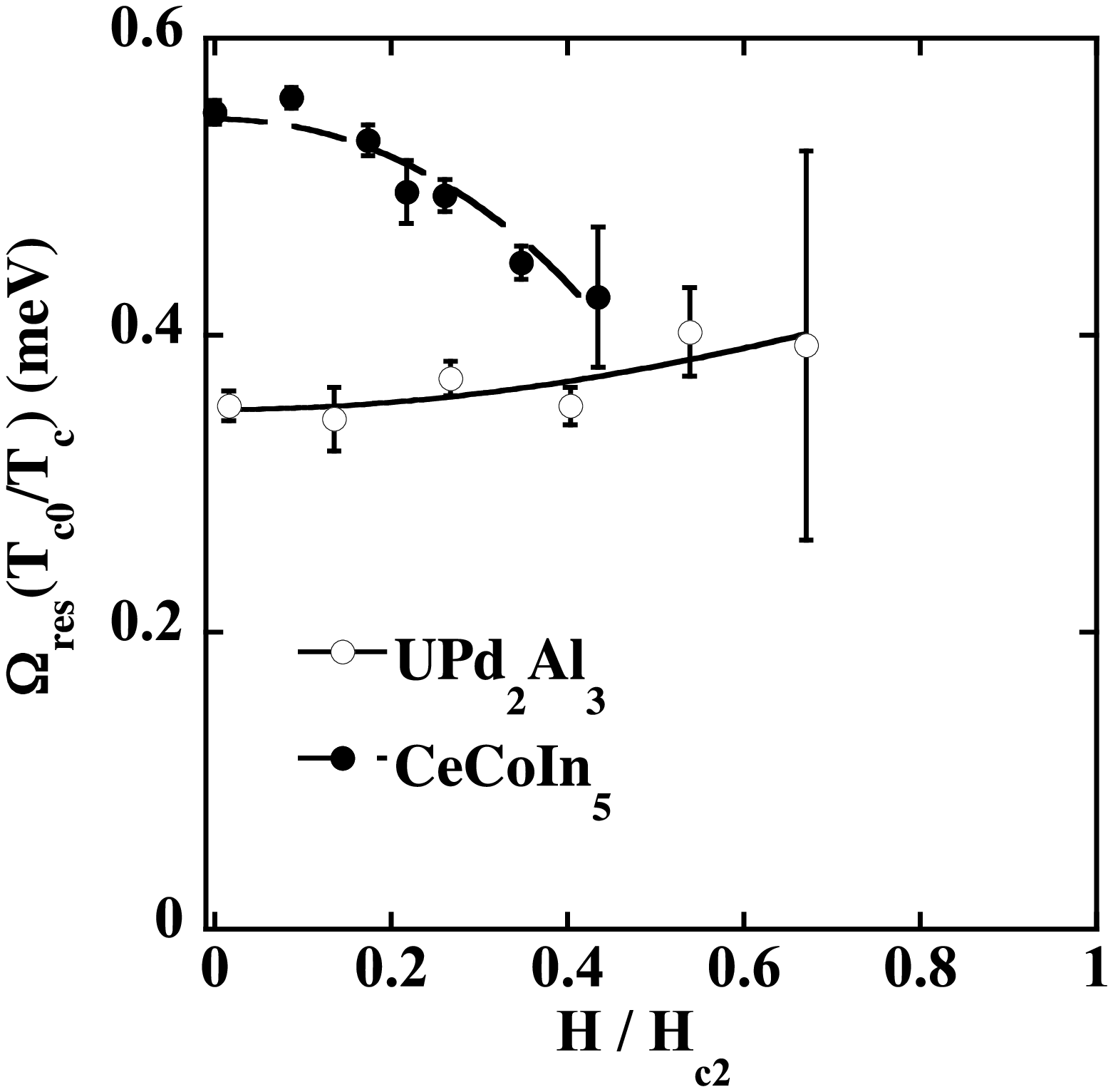}
%\vspace{ -2cm}
\caption{$\Omega_{res} \times T_{c0}/T_{c}$ as a function of $H/H_{c2}$ for CeCoIn$_{5}$ ($H$ // [1, $\bar{1}$, 0] \cite{Panarin1}) and UPd$_{2}$Al$_{3}$ ($H$// $b$\cite{Blackburn2}).}
\end{figure}

\section{Conclusion}
To conclude, evidences are given on the decrease of the resonance peak energy of CeCoIn$_{5}$ when superconductivity is suppressed by magnetic field or La substitution. In both cases the lineshape substantially broadens.
The Zeeman effect on the resonance surprisingly corresponds to the 4$f$ moment of the crystal field ground state, which can be considered as the field action on the individual $S$=1/2 quasiparticles. The faster decrease of $\Omega_{res}(H)$ with respect to $T_{c}(H)$ is tentatively related to the proximity of the magnetic field induced antiferromagnetic phase and a possible soft mode behaviour.

\bigskip

We acknowledge V. Michal for usefull discussions.


\begin{thebibliography}{99}
\bibitem{Note} For a magnetic field of 4.2 T applied along the $b$-axis ($Hc_{2}$=3.3 T), the magnetic moments rotate in the hexagonal plane and align perpendicular to the field. In Figure 5, we consider only the data taken below 2.08 T for which this effect is negligible.
\end{thebibliography}
\end{document}